%% file: simon.tex
\documentclass[a4paper,11pt]{article}
\usepackage[english]{babel}
\usepackage[latin1]{inputenc}
\usepackage[T1]{fontenc}
\usepackage{amssymb}
\usepackage{amsmath}
\usepackage{vmargin}
\usepackage{stmaryrd}
\usepackage{graphicx}
\usepackage{color}

\newcommand{\dom}{\text{dom}}
\newcommand{\incl}{\subseteq}
\newcommand{\Prob}{\mathbb{P}}

\newcommand{\R}{{\mathbb R}}
\newcommand{\Z}{{\mathbb Z}}

\newcommand{\pa}[1]{\left(#1\right)}
\newcommand{\cro}[1]{\left[#1\right]}
\newcommand{\abs}[1]{\left|#1\right|}
\newcommand{\acco}[1]{\left\{#1\right\}}
\newcommand{\ket}[1]{\left|#1\right>}

\newcommand{\Sim}[1]{{\bf Sim}}

\newtheorem{thm}{Theorem}
\newtheorem{prop}{Proposition}
\newtheorem{lemme}{Lemma}

\title{A quantum lower bound\\ for the query complexity of Simon's problem}

\author{Pascal Koiran, Vincent Nesme and Natacha Portier\\
Laboratoire de l'Informatique du Parall\'elisme\\
Ecole Normale Sup\'erieure de Lyon\\
46, allée d'Italie\\
69364 Lyon Cedex 07, France.\\
 \{Pascal.Koiran, Vincent.Nesme, Natacha.Portier\}@ens-lyon.fr }

\begin{document}

\maketitle

\date

\begin{center}\begin{minipage}{8cm} {\small Simon in his FOCS'94 paper
      was the first to show an exponential gap between classical and
      quantum computation. The problem he dealt with is now part of a
      well-studied class of problems, the hidden subgroup problems. We
      study Simon's problem from the point of view of quantum query
      complexity and give here a first nontrivial lower bound 
on the query complexity of a hidden subgroup problem, 
namely Simon's problem. Our bound is optimal up to a constant factor.}
\end{minipage}\end{center}

\bigskip

\section{Introduction}

Given an Abelian group $G$ and a subgroup $H\leq G$, a function
$f:G\to X$ is said to be hiding $H$ if $f$ can be defined in a
one-to-one way on $G/H$. 
More precisely, $f$ hides $H$ if and only if 
$$\forall g,g'\in G\; \pa{f(g)=f(g')\iff \exists h\in H \;g=g'+h}$$
Suppose $G$ is a fixed group and $f$ is computed by an oracle: a
quantum black-box. We are interested here in algorithms that find the
hidden subgroup $H$. 
A large amount of documentation about the hidden subgroup problem 
can
be found in the book of Nielsen and Chuang \cite{NC00}\footnote{History of the
problem on page 246 and expression of many problems 
(order-finding, dicrete logarithm...) 
in terms of hidden subgroup problems on page 241.}.
Among all work already done about such algorithms one can cite Shor's
famous factoring algorithm \cite{Sh97}: 
it uses a period-finding algorithm, 
which is a special case of a hidden subgroup problem. 
In recent years, attention has shifted to non-Abelian hidden subgroup
problems but we will restrict our attention here to Abelian groups,
and in fact to the family of groups $\pa{\Z/2\Z}^n$.

In general, two kinds of complexity measures for black-box problems 
can be distinguished: 
query complexity, i.e., the number of times the function $f$ is
evaluated using the black-box, 
and computational or time complexity, i.e.,
 the number of elementary operations needed to solve the problem.
Typically, a hidden subgroup algorithm is considered efficient if its 
complexity (in query or in time, depending on the interest) is
polynomial in the logarithm of the cardinality of $G$.
For example, Kuperberg's
algorithm~\cite{K03} for the (non-Abelian) 
dihedral hidden subgroup problem
is subexponential (but superpolynomial) 
in both time and query complexities. 
We give here a first nontrivial lower bound on the query complexity 
of a hidden subgroup problem, namely, Simon's problem.

This problem is defined as follows: we are given a function 
$f$ from $G=\pa{\Z/2\Z}^n$ to a known set $X$ of size $2^n$, 
and we are guaranteed that the function fulfills Simon's promises, that is either:
\begin{itemize}
\item[(1)] $f$ is one-to-one, or
\item[(2)] $\exists s \neq 0\;\forall w,w'\; 
f(w)=f(w')\iff \pa{w=w' \vee w=w'+s}$.
\end{itemize}

The problem is to decide whether (1) or (2) holds. Note that (1) is
equivalent to ``$f$ hides the trivial subgroup $H=\{(0, \dots, 0)\}$''
and (2) is equivalent to ``$f$ hides a subgroup $H=\{(0, \dots, 0),
s\}$ of order 2''. The original problem \cite{S97} was to compute $s$
and the problem considered here is the associated decision problem.
Of course, any lower bound on this problem will imply the same one on
Simon's original problem. 
In his article, Simon shows that his problem can be solved by a
quantum algorithm which makes $O(n)$ queries in the worst case
and has a bounded probability of error.
The time complexity of his algorithm is linear in the time 
required to solve an $n \times n$
system of linear equations over $\pa{\Z/2\Z}^n$.
He also shows that any classical (probabilistic) algorithm for his
problem must have exponential query complexity.
In this paper we shall give a $\Omega(n)$ lower bound on the
query complexity of Simon's problem, thus showing that Simon's
algorithm is optimal in this respect.
As a side remark, note that Simon also gives a Las Vegas version of
his algorithm with expected query complexity $O(n)$.
Even better, Brassard and H{\o}yer~\cite{BH97} have given an 
``exact polynomial time'' quantum algorithm for Simon's problem 
(i.e., their algorithm has a polynomial worst case running time and 
zero probability of error).

The two main methods for proving query complexity lower bounds in
quantum computing are the adversary method of Ambainis and the
polynomial method 
 (for an excellent review of these methods in French, read \cite{PP}).
We shall use the polynomial method, which was introduced in quantum
complexity theory in \cite{BBCMW01}.
There are recent interesting applications of this method to the
collision and element distinctness problem~\cite{AS04,Kutin03}. 
All previous applications of the polynomial method ultimately rely on
approximation theory lemmas of Paturi~\cite{P92} or Nisan and Szegedy~\cite{NS94}.
Besides the application to a new type of problems (namely, the hidden
subgroup problems) we also contribute to the development of the method
by applying it in a situation where these lemmas are not applicable.
Instead, we use an apparently new (and elementary) approximation theory result:
Lemma~\ref{degree} from section~\ref{lowerbound}.

In future work we plan to apply the polynomial method to other hidden
subgroup problems.
For a start, it seems that the groups $(\Z/p\Z)^n$ where $p$ is a fixed prime
can be handled in essentially the same way.

\section{Preliminaries}

We assume here that the reader is familiar with the basic notions 
of quantum computing \cite{NC00,H01} and we now present the polynomial method.
Let $A$ be a quantum algorithm solving Simon's decision problem. 
Without loss of generality, we can suppose that for every $n$ the
algorithm $A$ acts like a succession of operations
$$U_0,  O, U_1, O ,\dots , O , U_{T(n)}, M$$
on a $m$-qubit, for some $m\geq 2n$, starting from state $\ket{0}^{\otimes m}$. 
The $U_i$ are unitary operations and $O$ is the call to the black-box function:
if $x$ and $y$ are elements of $\{0,1\}^n$ then $O\ket{x,y,z}=\ket{x,y\oplus f(x),z}$.
The operation $M$ is the measure of the last qubit.
There are some states of $(m-1)$-qubits $\ket{\phi_0(f,n)}$ and
$\ket{\phi_1(f,n)}$ (of norm possibly less than 1) such
that 
$$U_{T(n)} O U_{T(n)-1} O \dots O U_{0}\ket{0}^{\otimes m}=\ket{\phi_0(n,f)}\otimes \ket{0}+\ket{\phi_1(n,f)}\otimes\ket{1}.$$
After the measure $M$, the result is $0$ (reject) with probability $||\phi_0(n,f)||^2$ and
$1$ (accept) with probability $||\phi_1(n,f)||^2$. 
The algorithm $A$ is said to solve Simon's problem
with bounded error probability $\epsilon$ 
if it accepts any bijection with probability at least 
$1-\epsilon$ and rejects every other function fullfilling Simon's promise 
with probability at least $1-\epsilon$. By definition, the query complexity
of $A$ is the function $T$.
Here is our main result.
\begin{thm}\label{bigthm}
If $A$ is an algorithm which solves Simon's problem
with bounded error probability $\epsilon$ and query complexity $T$,
then we have $T(n)\geq \frac{n+2+\log_2(2-4\epsilon)}{8}$
for every large enough integer $n$.
\end{thm}
As explained in the introduction, our proof of this theorem is based
on the polynomial method.
Lemma~\ref{polynomial} below is the key observation on which this
method relies.
We state it using the formalism of \cite{AS04}:
if $s$ is a partial function from $\pa{\Z/2\Z}^n$ to $E$ and $f$ a function 
from $\pa{\Z/2\Z}^n$ to $E$, $|\dom(s)|$ denotes the size of the domain of
$s$. Moreover, we define:
$$I_s(f)=\left\{\begin{array}{cl}
1 & \text{if $f$ extends $s$} \\
0 & \text{otherwise.}
\end{array}\right.$$ 

\begin{lemme} \label{polynomial} \cite{BBCMW01}
If $A$ is an algorithm 
of query complexity $T$,
there is a set $S$ of partial functions from $\pa{\Z/2\Z}^n\to E$
such that for all functions $f:\pa{\Z/2\Z}^n\to E$, 
$A$ accepts $f$ with probability 
$$P_n(f)=\sum\limits_{s\in S}\alpha_s I_s(f)$$ 
where for every $s\in S$ we have $|\dom(s)|\leq 2T(n)$ 
and $\alpha_s$ is a 
real number.
\end{lemme}

The goal is now to transform $P_n(f)$ into a low-degree polynomial 
of a single real variable. 
This is achieved in Proposition~\ref{upperbound}.
We can then prove and apply
our lower bound result on real polynomials (Lemma~\ref{degree}).

\section{Main proof} \label{lowerbound}

An algorithm for Simon's problem is only supposed to distinguish
between the trivial subgroup and a hidden subgroup of cardinality 2.
To establish our lower bound, we will nonetheless need to examine
its behavior  on a black-box hiding a subgroup 
of arbitrary order (a similar trick is used in~\cite{AS04} and~\cite{Kutin03}).
Note that this ``generalized Simon problem''
(finding an arbitrary hidden subgroup of $\pa{\Z/2\Z}^n$) 
can still be solved in $O(n)$ queries and bounded probability of error by
essentially the same algorithm, see for instance~\cite{H01}.

From now on we suppose that $A$ is an algorithm solving Simon's problem
with bounded error probability $\epsilon<\frac{1}{2}$ and query complexity $T$. Moreover,
$P_n(f)=\sum\limits_{s\in S}\alpha_s I_s(f)$ as given by lemma~1.

For $0\leq d\leq n$ and $D=2^d$, let $Q_n(D)$ be the probability that 
$A$ accepts $f$ when $f$ is chosen uniformly at random among the functions from $\pa{\Z/2\Z}^n$ 
to $E$ hiding a subgroup of $\pa{\Z/2\Z}^n$ of order $D$. Of course, $Q_n(D)$ is only
defined for some integer values of $D$ and it can be extended in many
different ways. By abuse of language
we will say that $Q_n$ is a polynomial of degree $d$ if it can be interpolated by a polynomial of degree $d$.

The point of this definition is that we have a bound on some values of
$Q_n$, and a gap between two of them. 
Namely, we have:

\begin{enumerate}

\item for any integer $d\in\cro{0;n}$, $0\leq Q_n(2^d)\leq 1$ (this
  number is a probability), and

\item $Q_n(1)\geq 1-\epsilon$ and $Q_n(2)\leq \epsilon$, 
hence  $|Q'_n(x_0)| \geq 1-2\epsilon > 0$ for some $x_0 \in\cro{1;2}$.
\end{enumerate}
If we denote by $X_D$ the set of functions 
hiding a subgroup of order $D$, 
by Lemma~\ref{polynomial} we have
$Q_n(D)=\sum\limits_{s\in S}
\pa{\frac{\alpha_s}{\abs{X_D}}\sum\limits_{f\in X_D} I_s(f)}$. 
Hence 
\begin{equation} \label{linear}
Q_n(D) = \sum_{s\in S} \alpha_s Q_n^s(D),
\end{equation}
where $Q_n^s(D)$ is the probability that a random function $f$ hiding 
a subgroup of order $D$ extends $s$.
We now prove that $Q_n$ is a low-degree polynomial.
By~(\ref{linear}), it suffices to bound the degree of $Q_n^s$.
Let us start by counting subgroups:

\begin{lemme} \label{subgroups}
Let $n$ and $k$ be nonnegative integers. \\
The group $\pa{\Z/2\Z}^n$ has exactly $\beta(n,k)=\prod\limits_{0\leq i<k}\frac{2^{n-i}-1}{2^{k-i}-1}$ distinct subgroups of order $2^k$.
\end{lemme}

\noindent {\bf Proof}\\
We look at $\pa{\Z/2\Z}^n$ as a vector space over the field $\Z/2\Z$: 
from this point of view the subgroups are the subspaces. 
We start by counting the number of free $k$-tuples of vectors. For the first $v_0$, 
we can choose anything but $0$, so there are $2^n-1$ choices. For the second vector $v_1$
we can choose anything but $0$ and $v_0$; $2^n-2$ possibilities remain. For the
third vector, any linear combinaison of $v_0$ and $v_1$ is forbidden: there are 4 of them.
In general, the number of free $k$-tuples of vectors 
is $\alpha(n,k)=\prod\limits_{0\leq i<k}\pa{2^n-2^i}$. 
Each subspace of dimension $k$ can be generated by $\alpha(k,k)$ 
different $k$-tuples, 
so the total number of subspaces of dimension~$k$ is
$\frac{\alpha(n,k)}{\alpha(k,k)}=\prod\limits_{0\leq
  i<k}\frac{2^{n-i}-1}{2^{k-i}-1}$. 
Note that this formula is correct even if $k>n$, in which case $\alpha(n,k)=0$.
\hfill$\boxempty$

\begin{prop} \label{upperbound}
The polynomial $Q_n$ is of degree at most $2T(n)$.
\end{prop}

\noindent {\bf Proof}\\
By~(\ref{linear}), it suffices to show that for all partial functions 
$s:\pa{\Z/2\Z}^n\to E$ 
such that $|\dom(s)|\leq 2T(n)$, the probability $Q_n^s(D)$ that a random function $f$ hiding 
a subgroup of order $D$ extends $s$ is a polynomial in $D$ of degree at most $2T(n)$. 
So, let $s$ be such a partial function. 
We will proceed in three steps: 
we first examine the case where $s$ is a constant function, 
then the case where $s$ is injective
and finally the general case.

Let us therefore suppose that $s$ is constant and note $\dom(s)=\acco{a_i/i=1\ldots k}$, with $k\leq 2T(n)$,
the $a_i$'s being of course all different. A function $f$ hiding a subgroup $H$ extends $s$ 
if and only if $\acco{a_i-a_1/i=1\ldots k}\incl H$ and $f(a_1)=s(a_1)$.
So $Q_n^s(D)=Q_{s'}(D)$ where $s'(x)=s(x-a_1)$. 
We will thus suppose without loss of generality that $a_1=0$. 
Since $E$, the possible range for $f$, is of size $2^n$, we have $Q_n^s(D)=\frac{\lambda}{2^n}$, 
where $\lambda$ is the proportion, among the subgroups of order $D$, of those containing $\dom(s)$.
 Let $H'$ be the subgroup generated by $\dom(s)$, and $D'=2^{d'}$ its order, $d'$ being the dimension 
of $H'$ as a vector space. The number of subgroups of order $D$ containing $H'$ is equal 
to the number of subgroups of order $\frac{D}{D'}$ of $\pa{\Z/2\Z}^n/H'$, which is isomorphic 
to $\pa{\Z/2\Z}^{n-d'}$; so there are $\beta(n-d',d-d')$ of them. We then have 
$Q_n^s(D)=\frac{1}{2^n}\frac{\beta(n-d',d-d')}{\beta(n,d)}=\frac{1}{2^n}\prod\limits_{0\leq i<d'}\frac{2^{d-i}-1}{2^{n-i}-1}$, which is a polynomial in $D$ of degree $d'<\abs{\dom(s)}\leq 2T(n)$.

Let us now suppose that $s$ is injective. We still note in the same way 
$\dom(s)=\acco{a_i/i=1\ldots k}$.
A function $f$ hiding a subgroup $H$ extends $s$ if and only if the 
$a_i$'s lie in distinct cosets
of $H$ and $f$ takes appropriate values on these cosets; so 
$Q_n^s(D)=\nu\lambda$,
where $\lambda$ is the probability for a subgroup $H$ of order $D$ to 
contain none of the
$a_i-a_j(i\neq j)$ and $\nu$ is the probability to extend $s$ 
for a function $h$ 
hiding a subgroup $H$
of order $D$ that does not contain any of the $a_i-a_j(i\neq j)$.
First we compute $\nu$. 
For each subgroup $H$ of order $D$ that does 
not contain any of the
$a_i-a_j(i\neq j)$ there are $(2^n)(2^n-1)\dots(2^n-n/D+1)$ possible 
functions $f$: choose a
different value for each coset of $H$.
Among these functions, the number of them extending $s$ is 
$(2^n-k)(2^n-k-1)\dots(2^n-n/D+1)$: choose
a value for each coset not containing any $a_i$.
So $\nu=\frac{\pa{2^n-k}!}{\pa{2^n}!}$.
The probability $\lambda$ is equal to $1-\mu$,
 where $\mu$  is the probability for a 
subgroup $H$ of
order $D$ to contain some $a_i-a_j$ for some $i\neq j$.

By the inclusion-exclusion formula, we can expand $\lambda$  as follows: 
$$\lambda =1-\cro{\begin{array}{cl}
&\sum\limits_{i\neq j}\Prob (a_i-a_j\in H)\\
-&\sum\limits_{\scriptsize\begin{array}{c}i_1\neq j_1\\i_2\neq j_2\\ \{i_1;j_1\}\neq \{i_2;j_2\} 
\end{array}}\Prob (a_{i_1}-a_{j_1}\in H \wedge a_{i_2}-a_{j_2}\in H)\\
+&\cdots \\
-&\cdots \\
\vdots & \\
+ &\Prob(\forall i\neq j \;a_i-a_j \in H)
\end{array}}$$

Our study of the first case above shows that each term in this sum 
is a polynomial in $D$ of degree less than $d'$, 
where the order of the subgroup generated by the $a_i-a_j$'s is
$2^{d'}$. Since $a_i-a_j$ is always in the subgroup generated by
$\dom(s)$, $d'\leq |\dom (s)| \leq 2T(n)$.

Finally, in the general case the partial function $s$ is defined by 
conditions of the form
$$\left\{ \begin{array}{c}
s(a_1^1)=s(a_2^1)=\dots=s(a^1_{k_1})=b_1 \\ 
s(a^2_1)=s(a^2_2)=\dots=s(a^2_{k_2})=b_2 \\
\vdots \\
s(a^l_1)=s(a^l_2)=\dots=s(a^l_{k_l})=b_l
\end{array} \right.$$
with $b_1,\dots, b_l$ all different.
In the same way as before, we will suppose without loss of 
generality that $a_1^1=0$. Furthermore, since $f(a_i^j)=f(a_1^j)$ 
is equivalent to $f(a_i^j-a_1^j)=f(0)$ 
(i.e. $a_i^j$ and $a_1^j$ are in the same coset of $H$) 
we can remove each $a_i^j$, for $i,j>1$ from $\dom(s)$ 
and replace them by adding the point $a_i^j-a_1^j$ to $\dom(s)$ 
associated to the value $b_1$. The size of $\dom(s)$ does not increase.
It may happen that $s$ was already defined on one of these entries 
and that our new definition is contradictory. 
In that case there is simply no subgroup-hiding function $f$ extending $s$, 
so $Q_n^s$ is simply the null polynomial and we are done. 
We will therefore consider only conditions of the form:
$$\left\{ \begin{array}{c}
s(0)=s(a_2^1)=\dots=s(a^1_{k_1})=b_1 \\ 
s(a^2)=b_2 \\
\vdots \\
s(a^l)=b_l
\end{array} \right.$$
The probability $Q_n^s(D)$ that a function $f$ hiding a 
subgroup of dimension $D$ extends
$s$ is the probability $Q_1$ that $f$ satisfies 
$f(0)=f(a_2^1)=\dots=f(a^1_{k_1})=b_1$
times the probabilty $Q_2$ that $f$ extends $s$ 
given that $f(0)=f(a_2^1)=\dots=f(a^1_{k_1})=b_1$.
We have already computed the first probability:
this is the case where $s$ is constant.
Let $H'$ be the subgroup generated by the $a_i^1$'s and  
$D'=2^{d'}$ its order; 
then $Q_1=\frac{1}{2^n}\prod\limits_{0\leq i<d'}\frac{2^{d-i}-1}{2^{n-i}-1}$. 
Let us define $s'$ on $G/H'$ as the quotient of $s$ if it exists  
(if not, this means again that $Q_n^s$ is the null polynomial, and we are done).
If $f$ satisfies $f(0)=f(a_2^1)=\dots=f(a^1_{k_1})=b_1$ then we can 
define $f'$ on $G/H'$ as the quotient of $f$; 
the condition ``$f$ extends $s$ and hides a subgroup of order $D$''
is equivalent to ``$f'$ extends $s'$ and hides a subgroup of order $D/D'$''. Since
$s'$ is defined by the condition 
$s'(H')=b_1, s'(a^2+H')=b_2, \dots, s'(a^l+H')=b_l$ and
is injective, our study of the second case shows that $Q_2=Q_{s'}(D/D')$ is a 
polynomial in $D$ of degree less than $|\dom(s')|$. 
Hence, $Q_n^s(D)$ is a polynomial in $D$ of degree at most 
$d'+|\dom(s')|\leq |\dom(s)|\leq 2T$.

\hfill$\boxempty$

Now that we have an upper bound on the degree of $Q$, 
let us find a lower bound. 
The following analogue of the lemmas of Paturi~\cite{P92} 
and Nisan-Szegedy~\cite{NS94} will help.
\begin{lemme}\label{degree}
Let $c>0$ be a constant and $P$ a polynomial with the following properties:
\begin{enumerate}
\item For any integer $0\leq i \leq n$ we have $\abs{P(2^i)}\leq 1$.
\item For some real number $1\leq x_0 \leq 2$ we have $\abs{P'(x_0)}\geq c$.
\end{enumerate}
Then $\deg(P)=\Omega\pa{n}$, and more precisely:
$\deg(P) \geq \min({n \over 2}, {n+2+\log_2 c \over 4}).$
\end{lemme}

\noindent {\bf Proof}\\
Let $d$ be the degree of $P$, and let us write
$P'(X)=\lambda\prod\limits_{i=1}^{d-1}(X-\alpha_i)$, where the
$\alpha_i$'s are real or complex numbers.
The polynomials $P'$ and $P''$ are respectively of degree $d-1$ and
$d-2$, so there exists an integer $a\in\cro{n-2d+2; n-1}$ such that
$P''$ has no real root in $\pa{2^a;2^{a+1}}$, and $P'$ has no root
whose real part is in this same interval.
If $d \geq n/2$ there is nothing to prove, so we may and we will
assume that $d\leq \frac{n}{2}$.
This implies in particular that $2^a\geq 4$.

The polynomial $P'$ is monotone on $\pa{2^a;2^{a+1}}$, for $P''$ has
no root in it. This means that $P$ is either convex or concave on this
interval, so that the graph of $P$ is either over or under its tangent
at the middle point of the interval, which is equal to
$\frac{2^a+2^{a+1}}{2}=\frac{3}{2}2^a$.
Suppose that $P'\pa{\frac{3}{2}2^a}$ is nonnegative (the case when it
is negative is similar). Then $P$ is increasing on $\pa{2^a;2^{a+1}}$,
since $P'$ has no root in this interval. Let $y=t(x)$ be the equation
of the tangent of $P$ at $\frac{3}{2}2^a$. If $t\pa{2^{a+1}}>1$, then
$P\pa{2^{a+1}}<t\pa{2^{a+1}}$, so $P$ is concave on
$\pa{2^a;2^{a+1}}$, hence $-1\leq P\pa{2^{a}}\leq t\pa{2^{a}}$.
But, since $P$ is monotone on $\pa{2^a;2^{a+1}}$,
$t\pa{\frac{3}{2}2^a}= P\pa{\frac{3}{2}2^a}\leq 1$.
Since $t(2^{a+1})-t\pa{\frac{3}{2}2^a}
= t\pa{\frac{3}{2}2^a} - t(2^{a})$,
it follows that $t\pa{2^{a+1}}\leq 3$ and $t\pa{2^{a+1}}-t\pa{2^a}\leq 4$.
The same inequality can also be derived if we assume $t\pa{2^{a}}<-1$,
and it is of course still true if $t\pa{2^{a}} \geq -1$ and
$t\pa{2^{a+1}} \leq 1$.
We conclude that that the inequality $t\pa{2^{a+1}}-t\pa{2^a}\leq 4$
always holds,
which implies that $0\leq P'\pa{\frac{3}{2}2^a}\leq
\frac{1}{2^{a-2}}$.
If we now include the case where $P'$ is negative,
we obtain the inequality
$$\abs{P'\pa{\frac{3}{2}2^a}}\leq \frac{1}{2^{a-2}}.$$

We therefore have 
\begin{equation} \label{quotient}
\abs{\frac{P'\pa{\frac{3}{2}2^a}}{P'(x_0)}}\leq \frac{1}{c
  2^{a-2}}\leq \frac{1}{c2^{n-2d}}.
\end{equation}

To conclude we need to state a simple geometric fact. Let $MBC$ be a triangle, $M'$ the orthogonal projection of $M$ onto $(BC)$, and $(d)$ the perpendicular bissector of $[BC]$. Let us suppose that $M$ is ``at the right of $(d)$'', i.e. $MC\leq MB$.

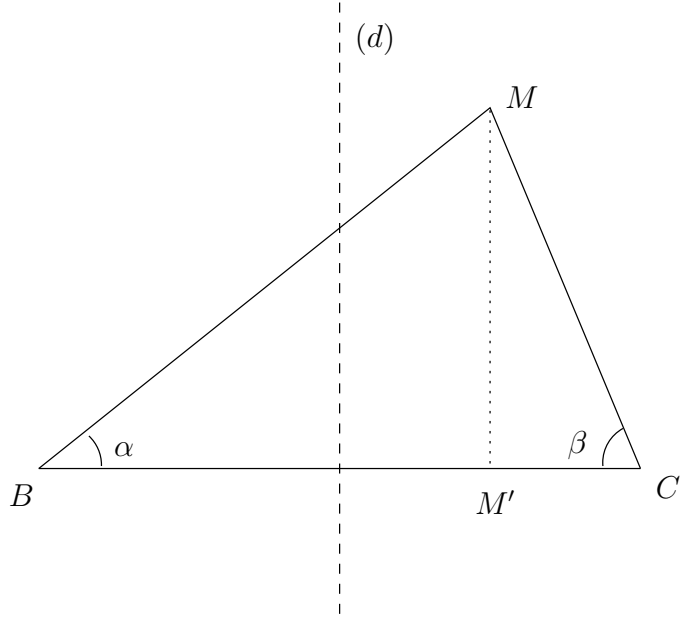
\begin{center}
\input{triangle.pstex_t}
\end{center}

Since $C$ is closer to the line $(MM')$ than $B$,
$\tan \alpha = MM'/BM' \leq \tan \beta = MM'/CM'$.
Hence $\alpha \leq \beta$, and $\cos \alpha \geq \cos \beta$,
i.e.: 
\begin{equation} \label{triangle}
\frac{MC}{MB}\geq \frac{M'C}{M'B}.
\end{equation}
Let $f:\pa{\begin{array}{rcl}
\R\setminus\acco{x_0}&\to&\R\\
x&\mapsto& \abs{\frac{\frac{3}{2}2^a- x}{x_0-x}}
\end{array}}$. Since $x_0<2^a<\frac{3}{2}2^a< 2^{a+1}$, a quick study
of this function shows that for all
$x\in\R\setminus\pa{\acco{x_0}\cup\pa{2^a;2^{a+1}}}$, $f(x)\geq \min
(1,f(2^a),f(2^{a+1})) \geq \frac{1}{4}$.

We will distinguish two cases for each $i\in\acco{1;\ldots;d-1}$.
\begin{enumerate}
\item If $\Re(\alpha_i)\leq \frac{1}{2}\pa{\frac{3}{2}2^a +x_0}$, then $\abs{\frac{\frac{3}{2}2^a- \alpha_i}{x_0-\alpha_i}}\geq 1$.
\item If $\Re(\alpha_i)> \frac{1}{2}\pa{\frac{3}{2}2^a +x_0}$, 
let us apply~(\ref{triangle}) to the points $M=\alpha_i$,
$M'=\Re(\alpha_i)$, $B=x_0$ and $C=\frac{3}{2}2^a$. 
We  obtain the inequality 
$$\abs{\frac{\frac{3}{2}2^a- \alpha_i}{x_0-\alpha_i}}\geq
\abs{\frac{\frac{3}{2}2^a- \Re(\alpha_i)}{x_0-\Re(\alpha_i)}}.$$
Remember though that no root of $P'$ has its real part in $\pa{2^a;2^{a+1}}$,
 so that $\abs{\frac{\frac{3}{2}2^a- \alpha_i}{x_0-\alpha_i}}\geq\frac{1}{4}$.
\end{enumerate}
We conclude that 
$\abs{\frac{\frac{3}{2}2^a- \alpha_i}{x_0-\alpha_i}}\geq\frac{1}{4}$
in both cases.
Taking~(\ref{quotient}) into account, 
we finally obtain the inequality $\frac{1}{4^{d-1}}\leq \frac{1}{c2^{n-2d}}$, 
hence $d\geq \frac{n+2+\log_2 c}{4}$.

\hfill$\boxempty$

We can now complete the proof of Theorem~\ref{bigthm}.
Let $A$ be our algorithm solving Simon's problem
with bounded error probability $\epsilon$ and query complexity $T$.
As pointed out before Lemma~\ref{subgroups}, 
the associated polynomial $Q_n$ satisfies $|Q'_n(x_0)| \geq 1-2\epsilon$ 
for some $\epsilon \in [1,2]$ and $Q_n(2^i) \in [0,1]$ for any $i \in
\{0,1,\ldots,n\}$. 
An application of Lemma~\ref{degree} to the polynomial $P=2Q_n-1$
therefore yields the inequality
$\deg(Q_n)\geq  \min\pa{{n \over 2}, {n+2+\log_2 \pa{2-4\epsilon} \over 4}}$.
Theorem~\ref{bigthm} follows since $\deg(Q_n)\leq 2T(n)$ by
Proposition~\ref{upperbound}.

\bibliographystyle{plain}
\bibliography{biblio}

\end{document}

%% file: triangle.pstex_t
\begin{picture}(0,0)%
\includegraphics{triangle.pstex}%
\end{picture}%
\setlength{\unitlength}{4144sp}%
\begingroup\makeatletter\ifx\SetFigFont\undefined%
\gdef\SetFigFont#1#2#3#4#5{%
  \reset@font\fontsize{#1}{#2pt}%
  \fontfamily{#3}\fontseries{#4}\fontshape{#5}%
  \selectfont}%
\fi\endgroup%
\begin{picture}(3900,3714)(1606,-3673)
\put(2251,-2686){\makebox(0,0)[lb]{\smash{{\SetFigFont{12}{14.4}{\rmdefault}{\mddefault}{\updefault}{\color[rgb]{0,0,0}$\alpha$}%
}}}}
\put(4966,-2671){\makebox(0,0)[lb]{\smash{{\SetFigFont{12}{14.4}{\rmdefault}{\mddefault}{\updefault}{\color[rgb]{0,0,0}$\beta$}%
}}}}
\put(5491,-2941){\makebox(0,0)[lb]{\smash{{\SetFigFont{12}{14.4}{\rmdefault}{\mddefault}{\updefault}{\color[rgb]{0,0,0}$C$}%
}}}}
\put(4411,-3031){\makebox(0,0)[lb]{\smash{{\SetFigFont{12}{14.4}{\rmdefault}{\mddefault}{\updefault}{\color[rgb]{0,0,0}$M'$}%
}}}}
\put(4591,-601){\makebox(0,0)[lb]{\smash{{\SetFigFont{12}{14.4}{\rmdefault}{\mddefault}{\updefault}{\color[rgb]{0,0,0}$M$}%
}}}}
\put(3691,-241){\makebox(0,0)[lb]{\smash{{\SetFigFont{12}{14.4}{\rmdefault}{\mddefault}{\updefault}{\color[rgb]{0,0,0}$(d)$}%
}}}}
\put(1621,-2986){\makebox(0,0)[lb]{\smash{{\SetFigFont{12}{14.4}{\rmdefault}{\mddefault}{\updefault}{\color[rgb]{0,0,0}$B$}%
}}}}
\end{picture}%